\documentclass[aps,showpacs,floatfix,twocolumn,superscriptaddress]{revtex4-1}
\usepackage{graphicx}
\usepackage{amssymb}
\usepackage{graphics, color}
\usepackage{amsmath}
\usepackage{subfigure}

\def\gsim{\, \rlap{$>$}{\lower 1.1ex\hbox{$\sim$}}\,}
\def\lsim{\, \rlap{$<$}{\lower 1.1ex\hbox{$\sim$}}\,}

\def\CO{{\cal O}}
\def\m{{\mu}}
 
 \def\n{{\nu}}
  \def\s{\sqrt}

\newcommand{\be}{\begin{equation}}
\newcommand{\ee}{\end{equation}}
 \newcommand{\bal}{\begin{align}}
 \newcommand{\eal}{\end{align}}
\newcommand{\ben}{\begin{equation*}}
\newcommand{\een}{\end{equation*}}
\newcommand{\bea}{\begin{eqnarray}}
\newcommand{\eea}{\end{eqnarray}}
\newcommand{\bean}{\begin{eqnarray*}}
\newcommand{\eean}{\end{eqnarray*}}
\newcommand{\bes}{\begin{subequations}}
\newcommand{\ees}{\end{subequations}}

\def\beq{\begin{equation}}
\def\eeq{\end{equation}}

\def\f {\frac}

\begin{document}
\vspace*{-1.5cm}
\begin{flushright}
  {\small
  MPP-2012-163\\
  }
\end{flushright}
\title{Normal modes and time evolution of a holographic superconductor after a quantum quench}
\author{Xin Gao}
\affiliation{Max-Planck-Institut f\"ur Physik (Werner-Heisenberg-Institut), \\
   F\"ohringer Ring 6,  80805 M\"unchen, Germany}
\affiliation{State Key Laboratory of Theoretical Physics, Institute of Theoretical Physics, Chinese
Academy of Sciences, P.O. Box 2735, Beijing 100190, China}
\author{Antonio M. Garc\'{\i}a-Garc\'{\i}a}
\affiliation{University of Cambridge, Cavendish Laboratory, JJ Thomson Avenue, Cambridge, CB3 0HE, UK}
\affiliation{CFIF, Instituto Superior T\'ecnico,
Universidade de Lisboa, Av. Rovisco Pais, 1049-001 Lisboa, Portugal}
\author{Hua Bi Zeng}
\affiliation{CFIF, Instituto Superior T\'ecnico,
Universidade de Lisboa, Av. Rovisco Pais, 1049-001 Lisboa, Portugal}
\affiliation{School of Mathematics and Physics, Bohai University
JinZhou 121000, China}
\author{Hai-Qing Zhang}
\affiliation{CFIF, Instituto Superior T\'ecnico,
Universidade de Lisboa, Av. Rovisco Pais, 1049-001 Lisboa, Portugal}

\begin{abstract}
We employ holographic techniques to investigate the dynamics of the order parameter of a strongly coupled superconductor after a perturbation that drives the system out of equilibrium. The gravity dual that we employ is the ${\rm AdS}_5$ Soliton background at zero temperature. We first analyze the normal modes
associated to the superconducting order parameter which are purely real since the background has no horizon.
We then study the full time evolution of the order parameter after a quench. For sufficiently a weak and slow perturbation we show that the order parameter undergoes simple undamped oscillations in time with a frequency that agrees with the lowest normal model computed previously. This is expected as the soliton background has no horizon and therefore, at least in the probe and large $N$ limits considered, the system will never return to equilibrium. For stronger and more abrupt perturbations higher normal modes are excited and the pattern of oscillations becomes increasingly intricate. We identify a range of parameters for which the time evolution of the order parameter become quasi chaotic. The details of the chaotic evolution depend on the type of perturbation used. Therefore it is plausible to expect that it is possible to engineer a perturbation that leads to the almost complete destruction of the oscillating pattern and consequently to quasi equilibration induced by superposition of modes with different frequencies.
\end{abstract}
\maketitle
%\section{Introduction}
%The description of the out of equilibrium dynamics of strongly
%interacting quantum systems remains one of the most challenging
%problems in theoretical physics.
Advances in cold atom physics and
numerical techniques have renewed the interest in the route and
conditions for thermalization in closed, strongly interacting, quantum systems
\cite{rigol,kinoshita} after a quench. Until recently theoretical progress had been
relatively slow as traditional analytical and numerical approaches are not in
general well suited to describe far from equilibrium dynamics.
However the application of the holographic principle, also called the AdS/CFT correspondence \cite{adscft}, has opened \cite{chesler,eke,murata,bizon,minwalla,taka1,sonner,starinets} new research avenues to tackle this problem.\\
This correspondence states that, under certain conditions, strongly coupled gauge conformal field theories (CFT) in $d$ dimensions are dual to a classical gravity theory in Anti de Sitter (AdS) space in $d+1$ dimensions.
%Information about the strongly coupled CFT can thus be obtained by simply %solving the equations of classical gravity with appropriate matter content and using the standard %dictionary among observables in the two sides of the duality.
%For instance,  a thermal state in the CFT is associated with a black hole of the same temperature in %the gravity dual. The quasinormal modes of %a AdS black hole correspond to the poles of the retarded %Green’s function of the dual conformal ﬁeld theory \cite{hubeny}.\\
%Within the range of applicability of the linear response approximation it has been possible to find explicit expressions for
%the transport properties such as the shear viscosity \cite{starinets} that are in qualitative agreement \cite{fopi} with quark gluon plasma %experimental results.
The out of equilibrium dynamics of the field theory has an especially
appealing interpretation in the gravity dual: it corresponds with
the time evolution of a mass distribution in an asymptotic AdS
background. Thermalization is thus related to the formation of a
black hole in the gravity dual \cite{bizon,minwalla}. In the context
of holographic superconductivity \cite{hhh} there are recent studies
on the time evolution of the order parameter
\cite{murata,sonner,silverstein,kam} at finite temperature. In
\cite{kam} it was found that, in the probe limit, normal modes (NM) of the order parameter have finite real and imaginary
parts (therefore usually called quasi normal modes) which indicates that, for long times, the superconducting order parameter
oscillates with an amplitude that decays exponentially as thermal
equilibrium is approached. Moreover Ref.\cite{kam} also identified the Goldstone mode that signals the
spontaneous breaking of the $U(1)$ symmetry at
sufficiently low temperature. These results have been recently
confirmed in \cite{sonner} by the explicit calculation of
the order parameter after a quench taking into account backreaction
effects. Similarly a previous study \cite{murata},
also including backreaction, showed that, for sufficiently long times after a
quench, deviations of the order parameter from the thermal equilibrium prediction are exponentially small. Moreover a quench in a Reissner-Nordstr\"om-AdS
black hole can induce an instability in the metric that leads to
holographic superconductivity \cite{murata}.
%Finally it was observed
%in \cite{silverstein} that a chemical potential that oscillates in
%time leads to enhancement of superconductivity.
%We note that in these papers the gravity background contains a blackhole, namely, the dual field theory was always at %finite temperature so, after a quench, the return to thermal equilibrium in a Boltzmann like fashion is expected.

Recent experimental \cite{kinoshita} and theoretical \cite{rigol,myself,leviemi} results in condensed matter points to a richer phenomenology. Integrability and localization  \cite{rigol,kinoshita,myself} can prevent, or slow down \cite{myself}, thermalization after a quench at zero temperature. In the context of weakly coupled superconductors \cite{leviemi} undamped oscillations of the order parameter after a quench have been observed provided that the final coupling is much larger than the initial one.
%In the opposite limit no oscillations are observed and the order %parameter vanishes exponentially.
 For intermediate quenches the superconducting gap is oscillatory in time with an amplitude that decays towards equilibrium in a power-law fashion \cite{leviemi}.

Motivated by these results we explore the time evolution of holographic
superconductors after a perturbation. Our main aim is to study the time evolution of the order parameter in a gravity background with no horizon so relaxation to equilibrium is not expected even after long times. In condensed matter that behavior is typical of insulator and integrable systems but we refrain to pursue this analogy to the gravity dual that we investigate,
a superconducting AdS Soliton background \cite{wiho,taka2} which has
a well defined limit at zero temperature. This is a key departure
from previous holographic studies \cite{sonner,murata,kam} which were carried
out at finite temperature, namely, in an AdS black hole background.
%We note that in the condensed matter results mentioned above the initial temperature is always zero.
 In order to gain insight into the late time
dynamics of the order parameter we study first the NM \cite{hubeny} related to the order parameter \cite{irene}. The AdS Solition background the
NM's are always real which suggests that the order parameter oscillates
without any damping. A detailed analysis of the time evolution after
a quench confirms this prediction, namely, after a perturbation the strongly coupled superconductor does not return to equilibrium. The oscillating pattern strongly depends on the perturbation details. In some cases we observed quasi chaotic oscillations which suggests that some sort of quasi equilibration is possible for some perturbations  even if there is no horizon.\\ % The origin of this interesting
%phenomenon is unclear though
%gap \cite{ftdual} that mimics the physics of a Mott insulator.\\

\section{The model}
We study the time evolution of a holographic superconductor in an
$AdS_5$ Soliton \cite{taka2} gravity background. This background
\cite{wiho} is constructed from the usual $AdS_5$ geometry
by a double Wick rotation followed by compactification of one of
the spatial dimensions. In order to have a smooth geometry it is
necessary to impose periodicity $\chi\sim\chi+\f{\pi L}{r_0}$ in the
compactified dimension. The resulting geometry has no horizon, it
looks like a cigar with a tip at $r=r_0$. The most stable
configuration which satisfies this constraint, the so called AdS
Soliton \cite{wiho}, is fully specified by the metric,
\begin{align}
\label{bubblesol}
 ds^2 = L^2\f{dr^2}{f(r)} + r^2( -dt^2 + dx^2 + dy^2 ) +
 f(r)d\chi^2
 \end{align}
with $ f(r) = r^2 - \f{r_0^4}{r^2}$.
In this background we introduce a charged scalar field $\Psi$
minimally coupled to a Maxwell field, which leads to the following
 five-dimensional Einstein-Maxwell-scalar gravity theory,
\bea
 S = \int d^5x \s{-g} (R + \f{12}{L^2} - \f{1}{4}F^{\m\n}F_{\m\n} \label{soaction} \\
 - |\nabla_\m \Psi - iqA_\m \Psi|^2 - m^2|\Psi|^2 ). \nonumber
\eea
We refer to (\ref{soaction}), together with the metric
(\ref{bubblesol}), as the  AdS soliton holographic superconductor
\cite{taka1}. We note that this model, unlike the usual
AdS-Schwarzschild superconductor, has a well defined zero
temperature limit.
% an important ingredient in order to study the
%conditions for thermalization in a quantum interacting system.

The properties of this model, investigated in  \cite{taka2} in the
probe limit and in \cite{benho} including the backreaction of the
scalar on the metric, depend on the value of the chemical potential
$\mu$, which corresponds to the time component of
the gauge field at the boundary (see below). For $\mu < \mu_c$ the
conductivity in the linear response regime vanishes and therefore
the field theory resembles that of a Mott insulator \cite{taka2}.
However for $\mu > \mu_c$ the scalar condenses and the system
becomes a superconductor though the background is still an AdS
Soliton. It is therefore expected that some of the Mott insulator physics might still be at play in this regime. For sufficiently large $\mu \gg \mu_c$ a
confinement-deconfinement transition occurs for any finite
temperature but not at strictly zero temperature \cite{benho}.
Superconductivity survives this transition but the gravity background becomes an AdS black hole \cite{hhh}.\\
 We restrict ourselves to the probe limit in which the backreaction
of the gauge field and scalar on the metric is negligible. We are motivated by the fact that, putting aside the expected black hole formation for stronger quenches, which is far beyond the scope of the paper,
it seems that \cite{benho}
including backreaction will not modify substantially the results obtained in the probe limit \cite{taka2}. In any case this is an interesting problem for future research.
\section{The equations of motion (EOM)}
Since we are interested in the time-evolution we seek
solutions of the EOM that depend on both the temporal and holographic coordinates:
%\begin{align}
% A&= \phi(r,t)dt  \ , \qquad \Psi = \Psi(r,t)
%\end{align}\label{soaction}
$A = (A_t, A_r, 0, 0, 0)$, $\Psi = |\Psi|e^{i\theta}$
where $A_t, A_r,  |\Psi|$ and $\theta$ are functions of $t$ and $r$
\cite{hai1}. In the following, we will work with the
gauge-invariant quantities $M=A-d\theta$. Without losing generality
we set $L=r_0= q = 1$ and $m^2 = -15/4$ so that the time evolution
depends only on the chemical potential. For numerical purposes it is more convenient to make the following change of variables, $z =1/r$ and $|\psi| = |\Psi|/ r^{3/2}$ so that the boundary is at $z=0$.
The resulting EOM are:
%\begin{align}
% \left({\frac{25}{4}} z^2-\phi^2- i\phi_t \right)z\psi + (-2+6 z^4)\psi_z +  \nonumber  \\
%z(-1+z^4)\psi_{zz}-2 iz\phi \psi_t +z\psi_{tt} &= 0 \label{ems1} \\
%2 z^4 \phi\psi\psi^{*}+(1+3 z^4)\phi_z-z \phi_{zz}+z^5 \phi_{zz}+ \nonumber \\
%i z^4 \psi^* \psi_t -i z^4 \psi \psi^*_t &= 0  \label{ems2} \\
%z^3 \psi^*\psi_z - z^3\psi \psi^*_z+i \phi_{tz} &= 0 \label{ems3}
%\end{align}

\begin{align}
   0=&|\psi| \bigg(-\left(z^4-1\right) {M_z}^2-z^4
   {M_t}^2+{9 z^6}/{4}\bigg)+4 z^7 \partial_z|\psi|
   \nonumber\\&+\left(z^4-1\right) z^4 \partial_z^2|\psi| +z^4 \partial_t^2|\psi|, \label{eom1}\\
   0=& \left(z^4-1\right) z \partial_{t}\partial_z{M_z}+\left(z^4+3\right)
   \partial_t{M_z}+\left(3 z^6+z^2\right)
   \partial_z{M_t}\nonumber\\&+\left(z^4-1\right) z^3 \partial_z^2{M_t}+2
   z^4 {M_t} |\psi| ^2,\label{eom2}\\
   0=&\partial_t^2{M_z}+2 z {M_z} |\psi| ^2+z^2
   \partial_t\partial_z{M_t},\label{eom3}\\
   0=& -\left(z^4-1\right) z |\psi|\partial_z{M_z} +z^3 |\psi|\partial_t{M_t}  +2 z^3 {M_t}
   \partial_t|\psi| \nonumber\\&
   -2 {M_z}
   \bigg(z \left(z^4-1\right)\partial_z |\psi| +\left(z^4+1\right) |\psi|
   \bigg). \label{eom4}
\end{align}
We note that Eqs. \eqref{eom2}, \eqref{eom3} and \eqref{eom4} are not
independent since Eq.\eqref{eom4} can be obtained from Eqs.
\eqref{eom2} and \eqref{eom3}.\\
{\it Boundary conditions.-}\\
In order to solve the EOM above it is necessary to specify initial conditions ($t=0$) and
boundary conditions at the tip $z=1$ and at the boundary
$z=0$.
Near the boundary and with the definitions above,
%\begin{eqnarray}
% \psi &=& \psi_{1}(t) z^{-1} + \psi_{2}(t) + \dots  \label{gx1}
% ,\\
% \phi &=& \mu(t) - \rho z^2 + \dots \nonumber
%\end{eqnarray}
\begin{eqnarray}
 |\psi| &=& |\psi_{1}(t)|  + |\psi_{2}(t)|z + \mathcal{O}(z^2) \label{gx1}
 ,\nonumber\\
 M_t &=& \mu(t) - \rho(t) z^2 + \mathcal{O}(z^3), \nonumber\\
 M_z &=& \mathcal{O}(z^3)
\end{eqnarray}
where $\mu$ and $\rho$ stand for the chemical potential and the charge density of the dual field theory respectively.
We choose solutions for which $|\psi_1| \equiv 0, |\psi_2| \neq 0$.
Following the standard AdS/CFT dictionary \cite{hhh} the
condensation of the operator $\CO_2$, \be \langle \CO_2(t) \rangle = |\psi_2(t)|
\label{co2} \ee
 stands for the order parameter of the dual field theory which describes the time evolution of the strongly-coupled superconductor.
Both the gauge field and the scalar must not be singular at the tip. A simple choice is $|\psi| = a_1  + a_2(1-z) + \dots$, $M_t = a_3 +
a_4(1-z) + \dots$ and $M_z=a_5+a_6(1-z)+\dots$, where the $a_i$'s
depend only on time. In order to obtain the initial conditions, $t=0$, we solve the EOM in the static limit for $\mu(t=0)\equiv\mu_i$.\\
%More details can be found in
%Sec.\ref{sect:fulltime}.
%%%%%%%%%%%%%%%%%%%%%%%%%%%%%%%%%%%%%%%%%%%%%%%%%%%%%%%%%%%%%%%%%%
\begin{figure}
\includegraphics[trim=0cm 0cm 8cm 21cm, clip=true,scale=0.6]{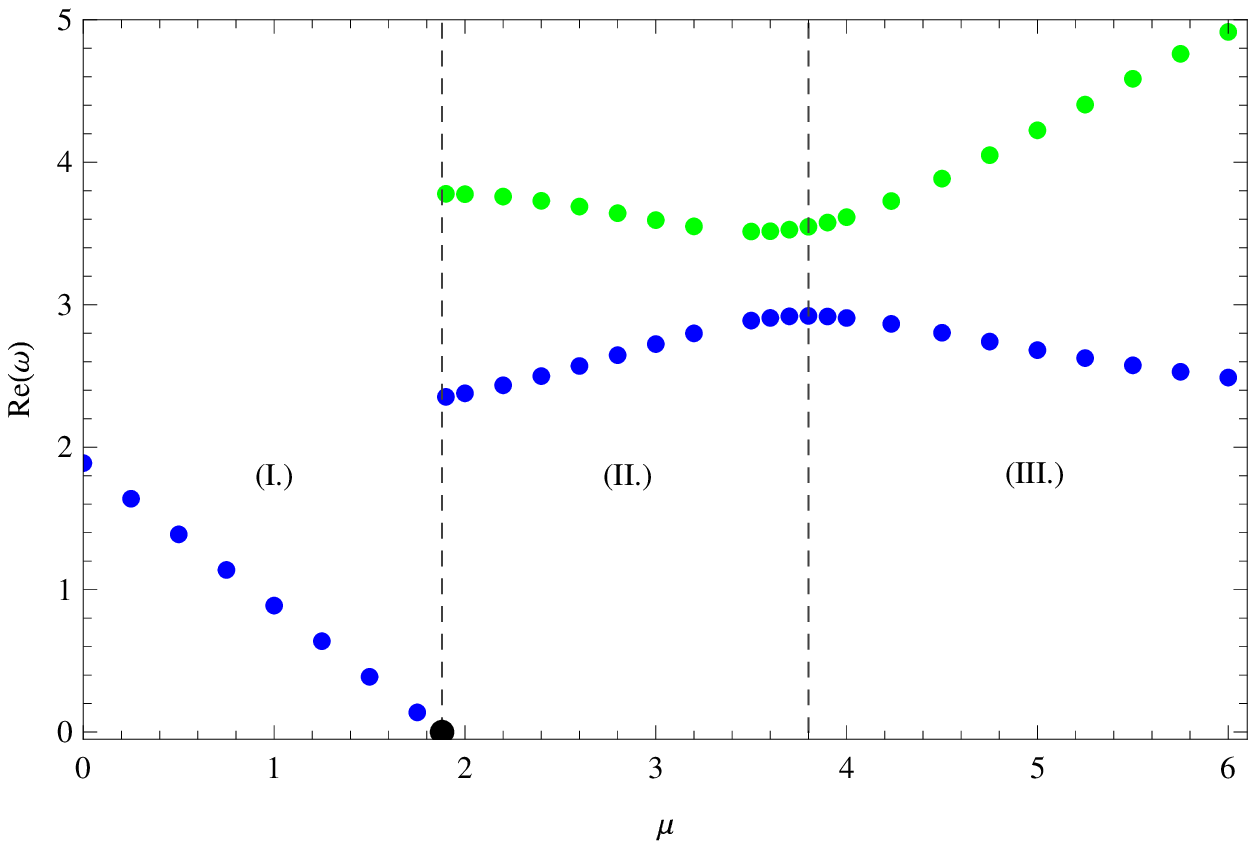}
\includegraphics[clip=true,scale=0.6]{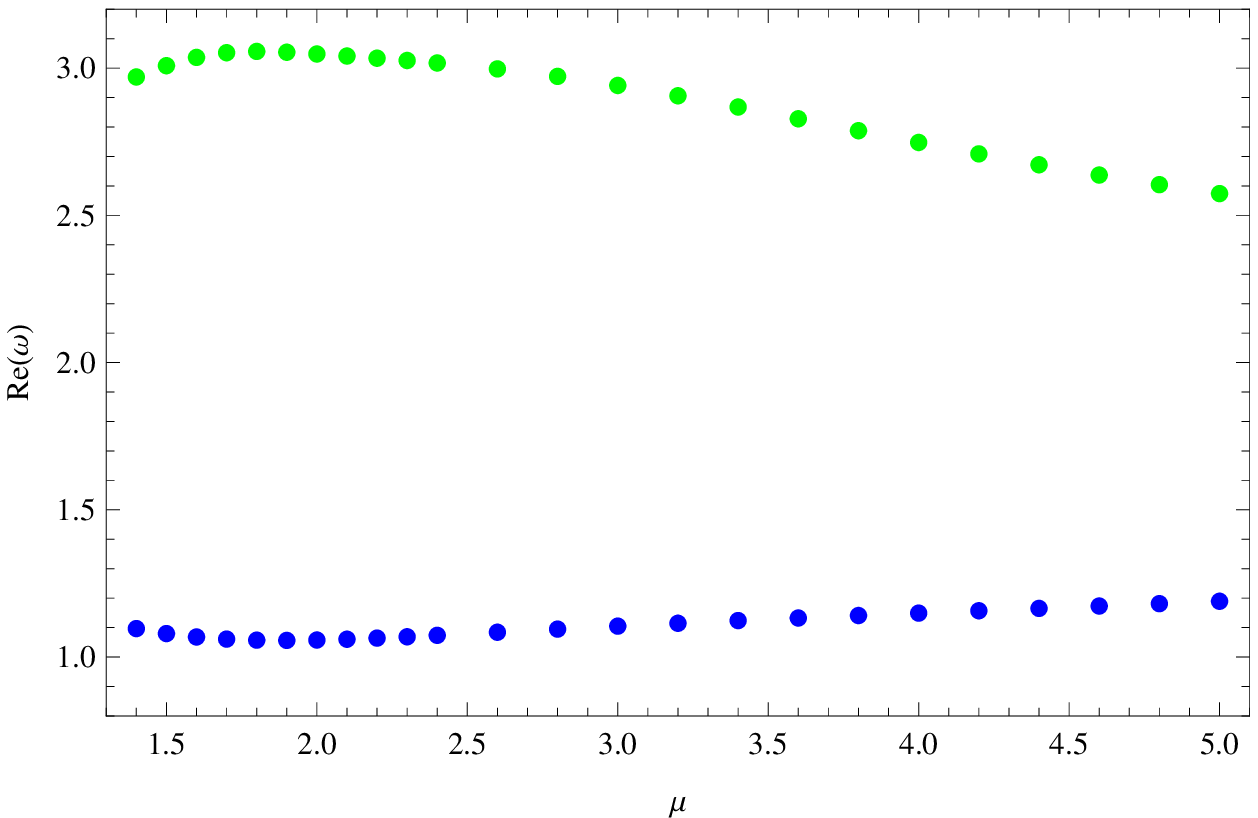}
\caption{(Color online) Upper: First (blue) and second (green) order NM $\mathrm{Re}[\omega]$ (\ref{omega}) as a function of $\mu$ for $\psi_1=0$ (no source).
% Physically this QNM corresponds to the lowest-lying pole of the  %Green's function associated to the scalar (\ref{gx1}).
A vanishing NM (black dot) signals the superconducting transition at $\mu = \mu_c \approx 1.88$ \cite{taka2}. Within the numerical precision of the calculation the imaginary part of the NM vanishes for any $\mu$. This suggests that after a quench the dual field theory does not return to equilibrium. The dynamics of the order parameter is expected to become more intricate for $\mu \sim 3.8$ since there is a minimum in the energy needed  to excite the second or higher order NM's.
Lower: First (blue) and second (green) order NM $\mathrm{Re}[\omega]$ (\ref{omega}) as a function of $\mu$ for $\psi_1 =1$. {The lowest modes are very close to the lowest frequency spectrum of the order parameter in the next section.}}
\label{fig1}
\end{figure}
%%%%%%%%%%%%%%%%%%%%%%%%%%%%%%%%%%%%%%%%%%%%%%%%%%%%%%%%%%%%%%%%%%%%%%%%%%%
\section{NM of the scalar field}
In order to gain insight on the asymptotic dynamics of $\langle \CO_2(t) \rangle$ we compute the NM associated to the scalar $|\psi|$. The NM's in the gravity theory correspond to the poles of the retarded Green's
function of the dual field theory \cite{hubeny}. Therefore it is
a powerful tool to investigate the asymptotic time evolution of the
order parameter after a small perturbation. We define $|\psi(t,z)|=|\psi_0(z) + e^{-i\omega
t}{\tilde \psi}(z)|$ where $\tilde \psi$ is small with respect to the static solution $\psi_0(z)$. We carry out a similar expansion for the other fields. We keep only terms in the EOM that are leading in
these small perturbations.
 The NM are defined as the discrete set,
 \begin{eqnarray}
 \omega(\mu) = \omega_R - i \omega_I \label{omega}
 \end{eqnarray}  frequencies  that solve the EOM. Our background has no horizon so it is expected these frequencies will be always real $\omega_I = 0$.
We used the determinant method \cite{kam} to calculate the NM. More specifically, we employ the Chebyshev differential matrix to convert the differentiation with respect to spacetime coordinate into a matrix.  The zeros of the determinant of the resulting matrix are the NM. In order to find the lowest mode, we calculated the lowest modes for different number of sites corresponding to the Chebyshev collocation grids in the radial direction. The physical lowest mode is the mode which is closest to zero and  is stable as the number of sites increases.

  A zero NM is a signature of a phase transition \cite{haiqing}. As was expected, see {the upper plot in} Fig. \ref{fig1}, it occurs at $\mu = \mu_c$ where the insulator-superconductor transition occurs \cite{taka2}. \\
 For $\mu<\mu_c$ (region I) , the EOMs decouple and the relevant NMs correspond to the pole of the retarded Green's function of the scalar. As shown in {the upper part in} Fig. \ref{fig1}, the NM's only contain a real part with $\omega_R \propto |\mu-\mu_c|$. This linear behavior of the NM can be predicted from inspection of the EOM's by noticing that $\omega(\mu) =\omega(0)-\mu$. \\ For $\mu > \mu_c$, in the superconducting phase, the calculation is more challenging since the scalar is coupled to a gauge field. In order to compute the NM's we use Chebyshev spectral method where the perturbations above are explicitly gauge-invariant. This is important in order to avoid spurious NM's. In {the upper part of} Fig. \ref{fig1} we plot the two lowest order NM's $\omega$ as a function of $\mu$. The black dot is the marginal stable mode which corresponds to $\omega=0$, this is the critical point.
%%%%%%%%%%%%%%%%%%%%%%%%%%%%%%%%%%%%%%%%%%%%%%%%%%%%%%%%%%%%%%%%%%
\begin{figure}[ht]
 \includegraphics[width=0.99\columnwidth,clip,angle=0]{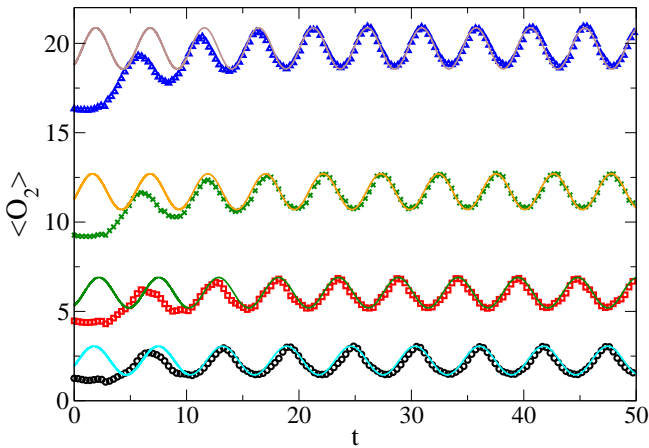}
 \caption[]{(Color online) $\langle \CO_2(t) \rangle$ (\ref{co2}) for a relatively weak quench $\psi_1(t)=J \tanh(vt)$ with $J=1$, $v=0.1$ and, from top to bottom, $\mu =5, 4,3$ and $2$. The solid lines are the best fit to a simple sinusoidal function {$\cos(\omega t)$} of frequency $\omega = 1.3,1.22.1.18,1.12$.}
 %   As $v$ increases the perturbation is more abrupt, higher energy modes are excited and the pattern of oscillations of the order parameter  $\langle \CO_2(t) \rangle$ become increasingly chaotic.}
 \label{fig2}
 \end{figure}
 %%%%%%%%%%%%%%%%%%%%%%%%%%%%%%%%%%%%%%%%%%%%%%%%%%%%%%%%%%%%%%%%%%
 As can be observed in {the upper part of} Fig. \ref{fig1} there is a jump of the NM at $\mu = \mu_c$.
According to \cite{taka2} this jump is directly related to the mass gap in the Soliton geometry, {which is corresponding to the mass gap in the phase transition from insulator to superconductor}. Region II is also characterized
by a growing (decreasing) lowest (second) NM's as $\mu$ increases.
This trend is reversed for $\mu \geq 3.8$ (region III). As was expected for all $\mu > \mu_c$,  $\omega_I = 0$ and $\omega_R \neq 0$ which indicates undamped oscillations of $\langle \CO_2(t) \rangle$.
 This is a striking indication that a holographic superconductor in a AdS Soliton background does not return to equilibrium after a perturbation.

{The results for the case in which the source $\psi_1=1$ is turned on, relevant for some of the quenches we study in the next section, are shown in the lower plot of Fig.\ref{fig1}.}
In the next section we shall see that for strong quenches the existence of several close NM's induces a qualitative change in the oscillating pattern of the order parameter, from simple sinusoidal to much more complex oscillations which eventually become quasichaotic. In order to fully confirm this quasichaotic behavior we study $\langle \CO_2(t) \rangle$ next.\\
\section{Full time evolution of $\langle \CO_2 \rangle$ after a quench}
%\label{sect:fulltime}
 NM's only provide information about the
dynamics for asymptotically long times and for small perturbations. In order to investigate shorter times and stronger perturbations we compute explicitly the full time evolution of $\langle \CO_2(t) \rangle$
(\ref{co2}) after a perturbation from equilibrium.\\

We study a perturbation in the order parameter
while keeping the chemical potential constant. Following Ref.\cite{sonner,basu}
we turn on the source $\psi_1(t)$ in the expansion of the scalar in the
boundary $\Psi(t,z)=\psi_1(t)z+\psi_2(t)z^2$. The quench details are determined by the functional form of $\psi_1(t)$.
We present results for \be \psi_1(t)=J\tanh(vt), \label{bas}\ee \cite{basu} and later in the paper \cite{sonner} for \be \psi_1(t)=\delta e^{-t^2/\tau^2.} \label{son}\ee
In order to solve the EOM's we use the spectral method for the
holographic dimension but, for technical reasons, we employ the Runge-Kutta
method in the time direction. Also for technical reasons we have added a small constant to the source $\psi_1(t)$ so it never vanishes.\cite{hai2}
%\footnote{
%The small constant is $\epsilon=0.01$ in $\psi_1(t)=\epsilon+J\tanh(vt)$. The reason for this small constant is that we have chosen Eq.\eqref{eom4} to solve %the time evolution of $M_t$ in the Runge-Kutta method. Therefore, in this case one needs to divide by $|\psi|$ on both sides of Eq.\eqref{eom4} in order to %satisfy the form of the Runge-Kutta algorithm. However, in this case an individual term $|\psi|$ will appear in the denominator. In order to avoid the %numerical divergence of the quench  $\psi_1(t)=J\tanh(vt)$ at $t=0$, one needs to add a small constant term in this quench. In fact one cannot avoid to use %Eq.\eqref{eom4} when solving the EoMs by Runge-Kutta method in time direction. Therefore, this is an indispensable trick to obtain meaningful results.}

The parameter $J$ control the strength of the
perturbation and $v$ its abruptness. For sufficiently small $v$ and $J$ the perturbation is slow and weak so only low energy excitations contribute to the dynamic of the order parameter. In the gravity dual these low energy excitations are nothing but the normal modes computed previously. Therefore we expect that, in this limit, only the lowest or a few NM contribute to the full time evolution. {In this case the order parameter should undergo simple oscillations with a frequency corresponding to the lowest NM.}
 %which is an increasing function of the chemical potential for $\mu < \mu_c$. For larger chemical potential the NM decreases with $\mu$.

We note that since our gravity background has no horizon, the NM's are purely real, so the order parameter does not return to
equilibrium after the quench. In Fig.\ref{fig2} we compare the order parameter dynamic in the $v \ll 1$ limit for different $\mu$ with a simple sinusoidal function. For $t >> 1/v$ the agreement is excellent.
%Also in agreement with the previous NM analysis the oscillation frequency has a (weakly) non monotonic dependence on $\mu$.
We have checked that numerical value of the condensate oscillation is in fair agreement with the lowest NM {in the lower part of Fig.\ref{fig1} around $\omega\sim1.1$}. Small disagreements might be possible because the quench in the full time calculation modifies the system, by turning on the source, in a way which is not necessarily captured by the linear NM analysis. Similar results (not shown) are also obtained for the quench Eq.(\ref{son}).
 %%%%%%%%%%%%%%%%%%%%%%%%%%%%%%%%%%%%%%%%%%%%%%%%%%%%%%%%%%%%%%%%%%
\begin{figure}[ht]
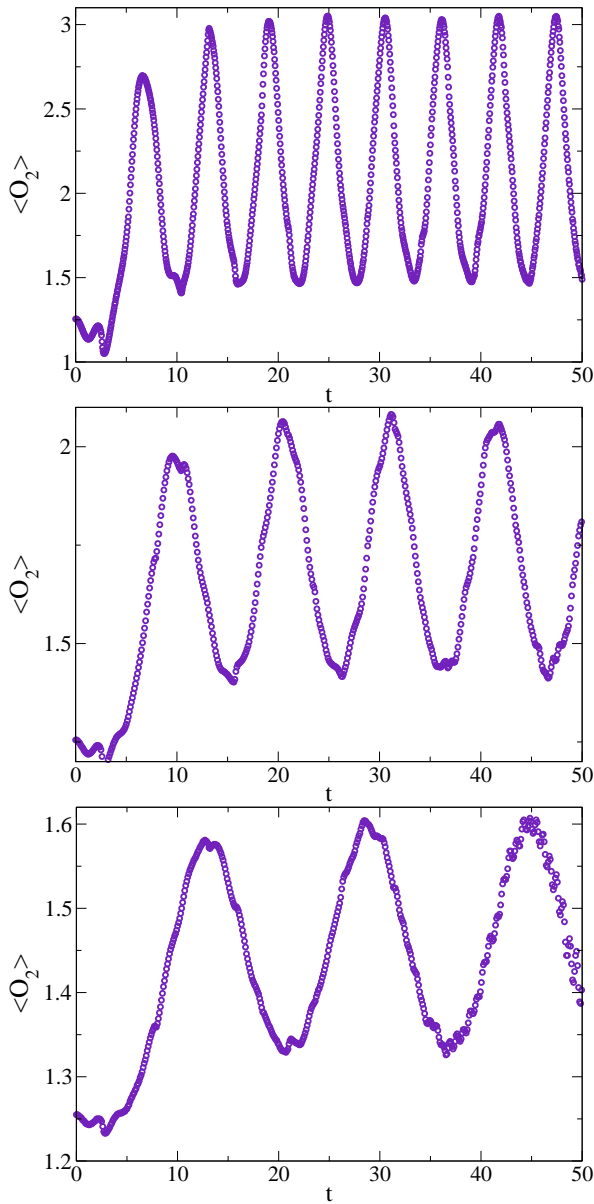

 \includegraphics[width=0.9\columnwidth,clip,angle=0]{m2v01J1.eps}
 \includegraphics[width=0.9\columnwidth,clip,angle=0]{m2v01J03.eps}
 \includegraphics[width=0.9\columnwidth,clip,angle=0]{m2v01J01.eps}
 \caption[]{(Color online) $\langle \CO_2(t) \rangle$ (\ref{co2}) for a quench $\psi_1(t)=J\tanh(vt)$ with $v=0.1$, $\mu=2$ and, from top to bottom, $J =1, 0.3$ and $0.1$. As $J$ increases the perturbation is stronger as the difference between the initial and final state increases as well. As a result, for sufficiently small $v$, the amplitude and frequency of $\langle \CO_2(t) \rangle$ become larger. Even for large $J$, assuming $v$ small, we did not observe any chaotic oscillations provided that $vJ \ll 1$. The frequency of the oscillations is still consistent with the lowest stable NM.}
 \label{fig3}
 \end{figure}
 %%%%%%%%%%%%%%%%%%%%%%%%%%%%%%%%%%%%%%%%%%%%%%%%%%%%%%%%%%%%%%%%%%

Still in the region $v \ll 1$ we now discuss the role of the perturbation strength $J$ in the time evolution of the order parameter.
 For $t \to 0$, the order parameter is not affected by the source $\langle O_2 \rangle = \Delta_0$. For $t \to \infty$, $\psi_1(t = \infty) \approx J$. The order parameter in this limit $\langle O_2 \rangle = \Delta_f$ is given by the static solution of the EOM's with $\psi_1=J$. It is clear \cite{sonner} that $\Delta_f \neq \Delta_i$. At least for small quenches, where only one mode is excited, we expect that the time evolution of the order parameter is closely related to these two parameters. More specifically we expect that the amplitude and frequency of the order parameter are an increasing function of $J$. In Fig.\ref{fig3} we show that the numerical calculation fully supports the theoretical prediction. We note that for a fixed small $v$ we did not observe a more complex oscillating pattern by only increasing $J$ provided that $vJ <1$. This is a consequence that for a sufficiently slow $v \ll 1$ quench the dynamics is almost adiabatic and not many higher energy modes are excited. Non-linear effects do not seem to be important in this regime as the frequency of the oscillations is still consistent with the lowest stable NM. { Please see the following TABLE.\ref{tab} which contains the NM's from the linear perturbation with $\psi_1=1, 0.3, 0.1$ compared to the frequencies of the order parameter from Fig.\ref{fig3}. We can see that the NM's and the frequencies are very close to each other.}
  \begin {table}[h]
 \caption{\label{tab} {The NM's from the linear perturbations compared to the frequencies from Fig.\ref{fig3}. } }
 \begin{eqnarray}
 \begin{tabular}{|c|c|c|c|}
   \hline
   % after \\: \hline or \cline{col1-col2} \cline{col3-col4} ...
   $J$ & NM's & Frequencies from Fig.\ref{fig3} & Errors \\
   \hline
   $1$ & $1.0577$ & $1.12$ & $5.56\%$ \\
   \hline
   $0.3$ & $0.6008$ & $0.61$ & $1.50\%$ \\
   \hline
   $0.1$ & $0.3887$ & $0.39$& $0.3\%$ \\
   \hline
    \end{tabular}
  \end{eqnarray}
    \end{table}

 %%%%%%%%%%%%%%%%%%%%%%%%%%%%%%%%%%%%%%%%%%%%%%%%%%%%%%%%%%%%%%%%%%
\begin{figure*}[ht]
\centering
 \includegraphics[width=0.9\columnwidth,clip,angle=0]{mu3v3J1.eps}
 \includegraphics[width=0.92\columnwidth,clip,angle=0]{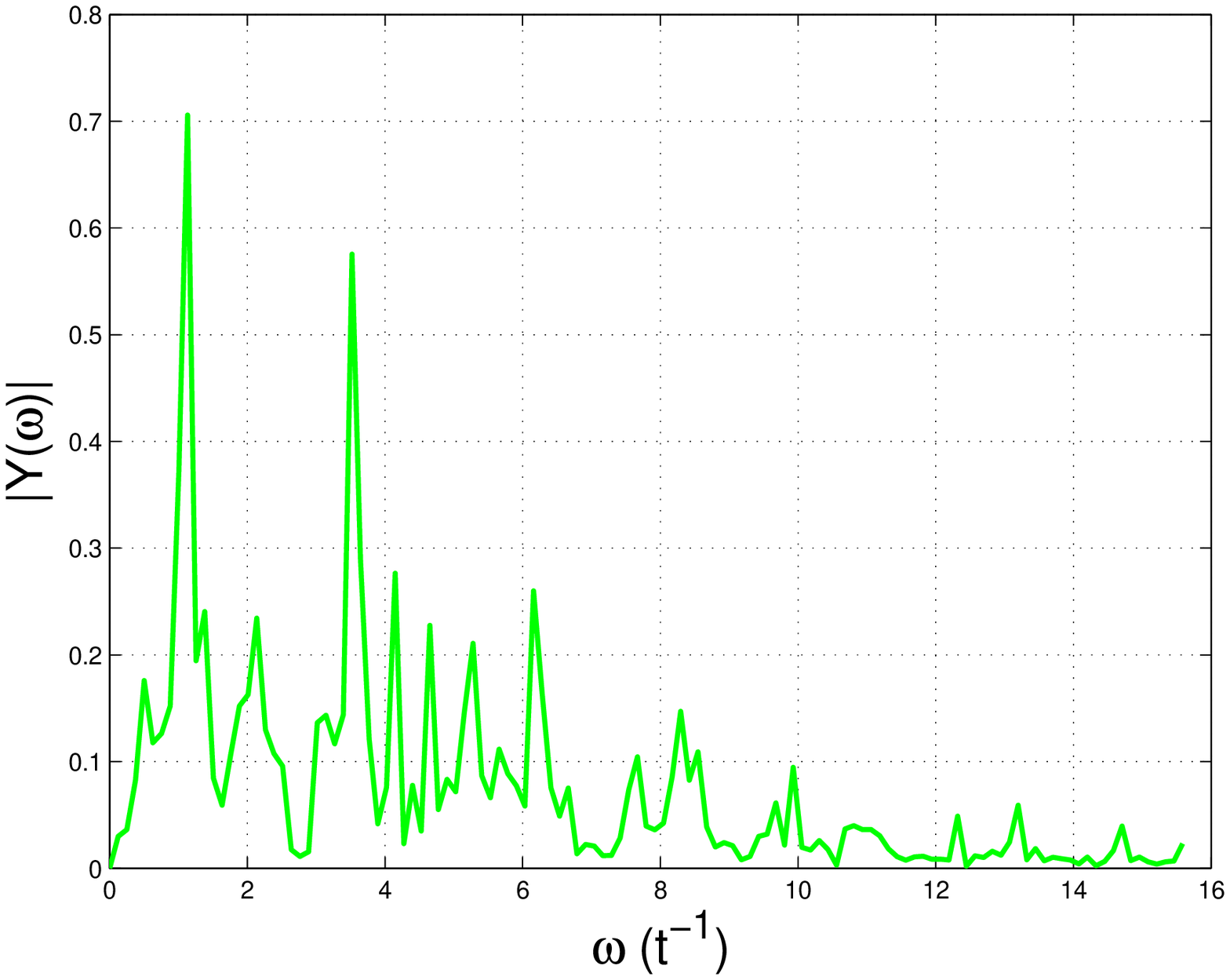}
 \includegraphics[width=0.9\columnwidth,clip,angle=0]{mu3v1J1.eps}
 \includegraphics[width=0.92\columnwidth,clip,angle=0]{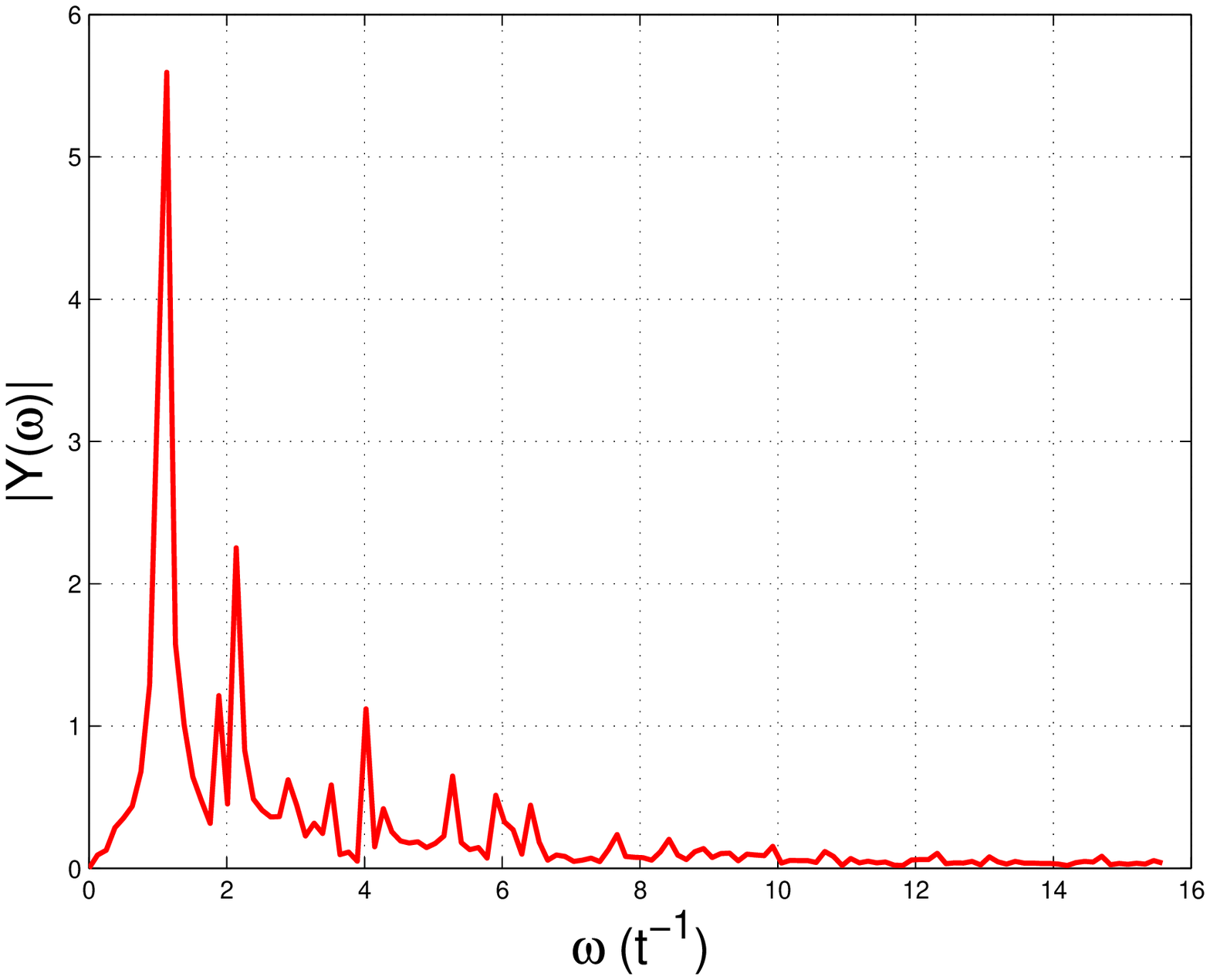}
 \includegraphics[width=0.9\columnwidth,clip,angle=0]{mu3v03J1.eps}
 \includegraphics[width=0.92\columnwidth,clip,angle=0]{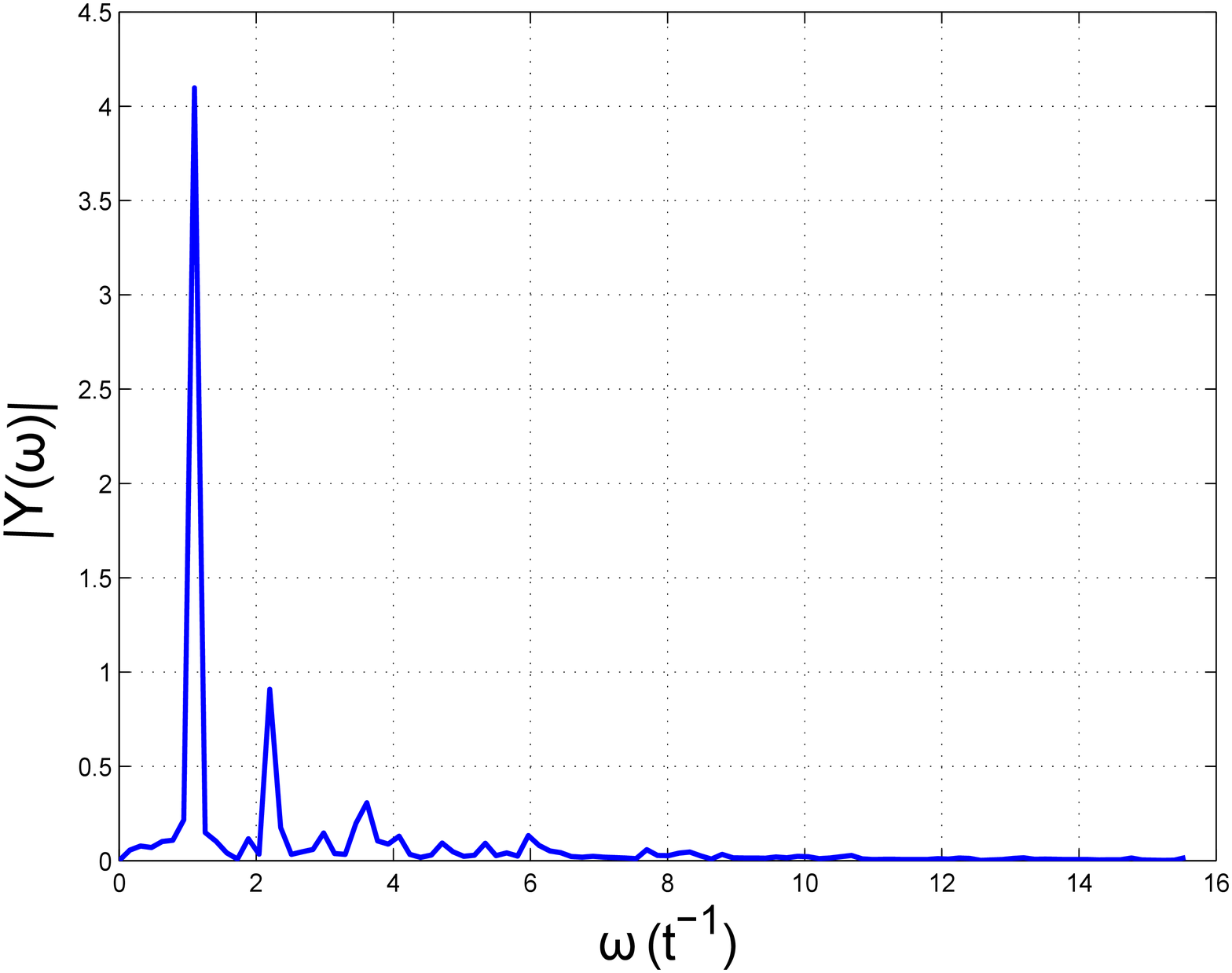}
 \caption[]{(Color online)Left panel: $\langle \CO_2(t) \rangle$ (\ref{co2}) for a quench $\psi_1(t)=J\tanh(vt)$ with $J=1$, $\mu=3$ and, from top to bottom, $v =3, 1$ and $0.3$. As $v$ increases the perturbation is more abrupt, higher energy modes are excited and the pattern of oscillations of the order parameter  $\langle \CO_2(t) \rangle$ become increasingly chaotic. Right panel: Fourier transform Eq.(\ref{FT}) of the order parameter time evolution in the left panel. From top to down, the first two highest mode correspond to $\omega=(1.1312, 3.5193), (1.1310, 2.1361), (1.0996, 2.1991)$, respectively. We note that for the quasi-chaotic spectra, the lowest mode is around $\omega=1.1$ which is still qualitatively consistent with the results depicted in Fig.\ref{fig2}. However, in this case many other higher modes also contribute to the time evolution of the order parameter, which results in a finite support for the Fourier transform. This is an indication that the motion is quasi-chaotic. Indeed as the quench becomes stronger, {\it i.e.}, $v$ increases, more higher modes will be excited as the time evolution gradually becomes more chaotic. }
 \label{fig4}
 \end{figure*}
 %%%%%%%%%%%%%%%%%%%%%%%%%%%%%%%%%%%%%%%%%%%%%%%%%%%%%%%%%%%%%%%%%%

 %%%%%%%%%%%%%%%%%%%%%%%%%%%%%%%%%%%%%%%%%%%%%%%%%%%%%%%%%%%%%%%%%%
\begin{figure*}[ht]
\centering
 \includegraphics[width=0.9\columnwidth,clip,angle=0]{mu3w1del05.eps}
 \includegraphics[width=0.92\columnwidth,clip,angle=0]{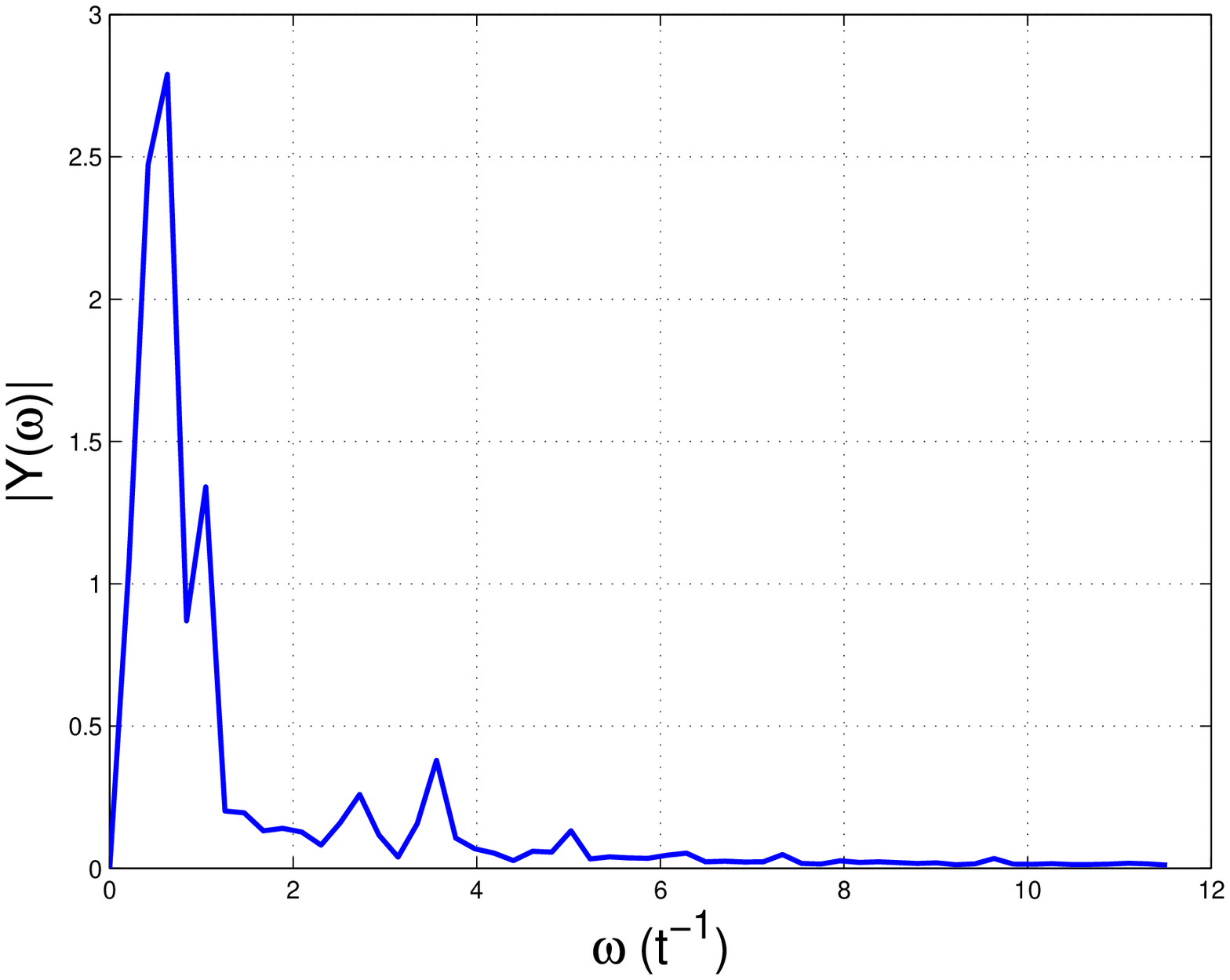}
 \includegraphics[width=0.9\columnwidth,clip,angle=0]{mu3w05del05.eps}
 \includegraphics[width=0.92\columnwidth,clip,angle=0]{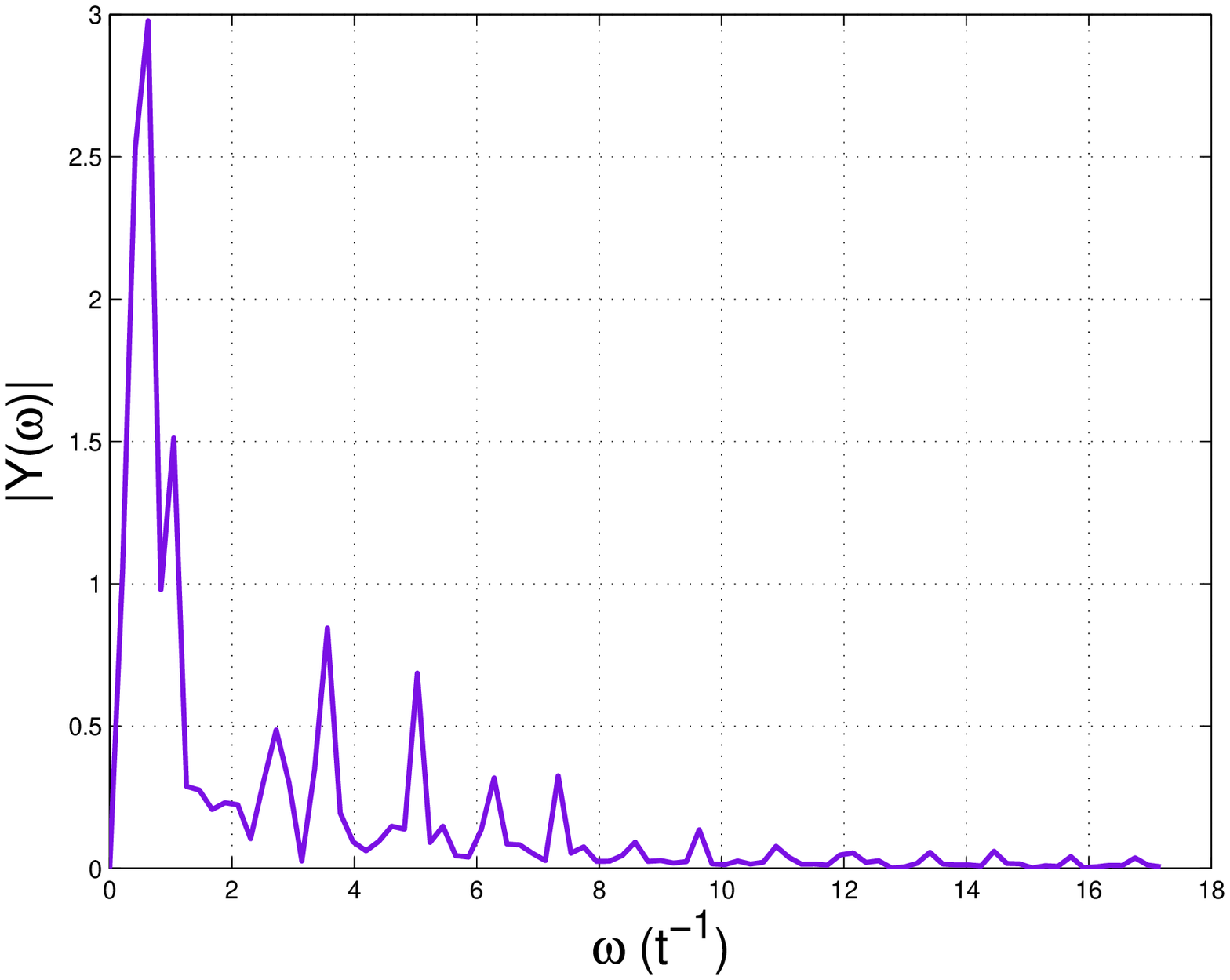}
 \includegraphics[width=0.9\columnwidth,clip,angle=0]{mu3w02del05.eps}
 \includegraphics[width=0.92\columnwidth,clip,angle=0]{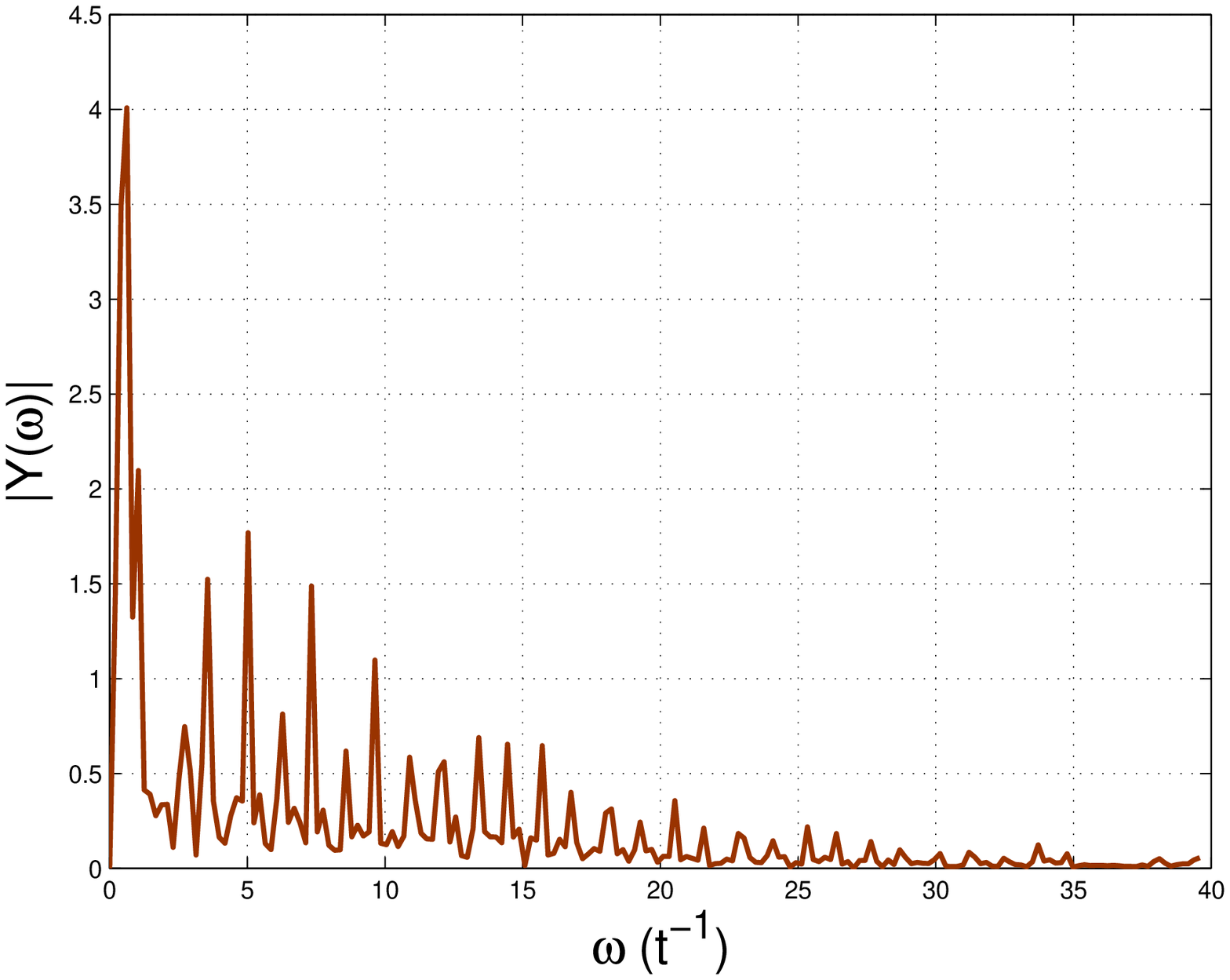}
 \caption[]{(Color online) Left panel: $\langle \CO_2(t) \rangle$ (\ref{co2}) for the quench Eq.(\ref{son}) with $\delta=0.5$, $\mu=3$ and, from top to bottom, $\tau =1, 0.5$ and $0.2$. As $\tau$ decreases the perturbation is more abrupt, higher energy modes are excited and the pattern of oscillations of the order parameter  $\langle \CO_2(t) \rangle$ becomes also increasingly chaotic but quantitatively different from the one obtained from the quench Eq.(\ref{bas}), see Fig.\ref{fig4}. Right panel: The Fourier transform Eq.(\ref{FT}) of the order parameter in the left panel. From top to down, the first two highest modes in all cases still correspond to $\omega=(0.6283,1.0472)$ for $\delta=0.5$, $\mu=3$. { That is expected as the position of the low energy mode should not depend on the quench strength.}  We can also see that when the quench becomes stronger, {\it i.e.}, $\tau$ becomes smaller, higher modes will be excited, the Fourier transform seems to have a finite support for small frequencies which suggests that the dynamic of the order parameter becomes more chaotic. }
 \label{fig5}
 \end{figure*}
 %%%%%%%%%%%%%%%%%%%%%%%%%%%%%%%%%%%%%%%%%%%%%%%%%%%%%%%%%%%%%%%%%%
We now turn to the case of a stronger perturbation.
 Strictly speaking, an abrupt perturbation, namely, a true quench,
corresponds to the limit $v \gg 1$ where the source $\psi_1$ reaches the asymptotic value $\approx J$ in a small time $t_Q \sim 1/v$ as compared with other time scales of the problem. Physically, abrupt
changes in the system induce high energy excitations. In the holographic
context this means that several NM's contribute simultaneously to the time
evolution of the order parameter. The linear approximation, in which the NM analysis is based on, breaks down so a full calculation of the time evolution of the order parameter is needed to have a good understanding of the dynamics. Qualitatively we expect the pattern of
oscillations of the order parameter to become more intricate as $v$ increases
and higher energy states are excited simultaneously. For sufficiently large
$v$, the superposition of many single modes should lead to a rather chaotic
pattern where single oscillations are not clearly distinguished. Non-linear effects should further enhance the chaoticity of the time evolution.  The results
depicted in Fig. \ref{fig4} fully confirm these theoretical predictions. We
note that for a larger value $\mu$, the chaotic region is observed for smaller
$v$'s. As was shown in the previous section more normal modes in the
region $\mu \sim 3.8$ are closer so it is expected that a less abrupt, smaller $v$,
quench will have a similar effect. Results for the quench Eq.(\ref{son}), see Fig. \ref{fig5}, show a similar qualitative behavior. However there are important quantitative differences. In the range of parameters that we can get meaningful results, the chaotic evolution is more evident for the quench Eq.(\ref{son}). While for the quench Eq.(\ref{bas}) the single oscillation induced for the lowest NM can still be distinguished for larger quenches ($v \gg 1$) though it is severely modified. By contrast to the other quench  Eq.(\ref{son}) no trace of the first normal mode remains, or it is easily observed, for larger quenches. We believe that the reasons for that behavior is that the quench Eq.(\ref{son}) is abrupter since the decay is Gaussian instead of exponential. Finally we note that oscillating patterns with a even richer structure, corresponding to smaller $v$, are harder to simulate numerically because the need of a smaller lattice spacing which is beyond our numerical capabilities.

 %%%%%%%%%%%%%%%%%%%%%%%%%%%%%%%%%%%%%%%%%%%%%%%%%%%%%%%%%%%%%%%%%%
\begin{figure}[ht]
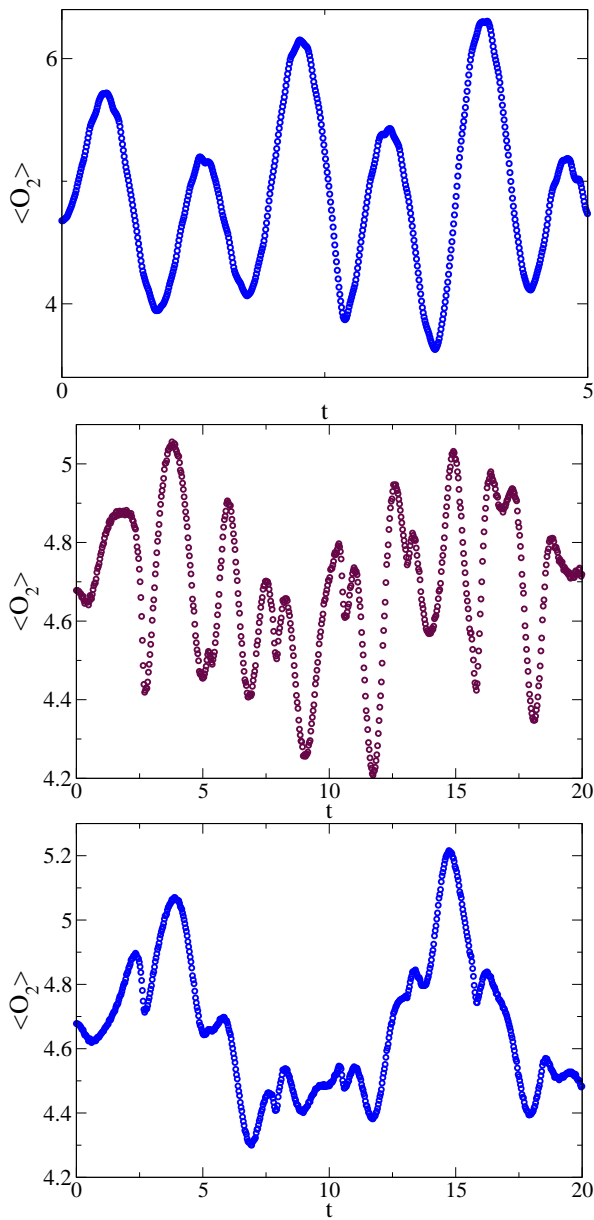

 \includegraphics[width=0.9\columnwidth,clip,angle=0]{mu3a01ome7.eps}
 \includegraphics[width=0.9\columnwidth,clip,angle=0]{mu3a01ome2.eps}
 \includegraphics[width=0.9\columnwidth,clip,angle=0]{mu3a01ome1_17.eps}
 \caption[]{(Color online) $\langle \CO_2(t) \rangle$ (\ref{co2}) after the perturbation Eq.(\ref{sina}) with $a=0.1$, $\mu=3$ and, from top to bottom, $\omega =7, 2$ and $1.17$. As $\omega$ increases the perturbation is theoretically more abrupt however $\langle \CO_2(t) \rangle$ becomes more chaotic in the opposite limit. We believe that this is due to the fact that in this region the dynamic is becoming controlled by the external frequency rather than by the natural NM associated to the order parameter.}
 \label{fig6}
 \end{figure}
 %%%%%%%%%%%%%%%%%%%%%%%%%%%%%%%%%%%%%%%%%%%%%%%%%%%%%%%%%%%%%%%%%%
We speculate that for a certain quench protocol it might occur that interference among the different NM excited by the quench, together with non-linear effects beyond the NM analysis, might lead to a dramatic destructive interference. Very likely the resulting order parameter evolution might be not very different from a weak form of equilibration induced by interference instead  of a finite temperature. In order to further explore this idea we also investigate the time evolution after a oscillatory perturbation, \be \psi_1(t)=a\cos(\omega t), \label{sina}\ee  with $a$ and $\omega$ real numbers. Results are shown in Fig.\ref{fig6}. For small frequencies around the smallest NM, $\omega \sim 1.17$, and for $\mu=3$, we have observed a relatively  small perturbation of the order parameter with a very intricate oscillatory pattern which, at least superficially, is not related to the original perturbation. However as the frequency of the perturbation increases, see Fig.\ref{fig6} the dynamic of the order parameter becomes more regular and closer to the perturbation. For $\omega=7$ the oscillating pattern seems to be a superposition of two sinusoidal functions with different frequency. We do not have a clear understanding of this phenomenon but it might be that weaker perturbation couple better to the normal modes that the stronger one.
\emph{
As a consequence many normal modes are more excited for weaker quenches than for stronger quenches.} More work is needed to fully solve this issue.

\section{Fourier Transformation of the time evolution}
In order to get a more quantitative description of the time evolution of the order parameter $\langle O_2(t)\rangle$, especially in the chaotic region, we have computed the Fourier transform of $\langle O_2(t)\rangle$,
\be\label{FT}
Y[\langle O_2\rangle](\omega)=A \int^{\infty}_{-\infty}\langle O_2(t)\rangle e^{-i\omega t}dt,
\ee
{ Where $A$ is a scaling constant which are different for each plots in Fig. \ref{fig4} and \ref{fig5}. However, $A$ is not important to the frequency positions in these plots, therefore, we did not show the values of $A$ explicitly in this paper.} Simple sinusoidal oscillations in time corresponds to a single peak at $\omega \neq 0$ in the Fourier spectrum. A more complex oscillatory pattern in time corresponds to several peaks being excited in the Fourier spectrum. Finally, a fully chaotic evolution in time leads to a Fourier spectrum with finite support in a large region of frequencies, due to overlapping of close Fourier modes, and with many peaks still superimposed over the finite background.

In the right panels of Figs.\ref{fig4} and Figs.\ref{fig5} we depict Eq.(\ref{FT}) corresponding to the time evolution for the parameters after quenches, Eq. (\ref{bas}) and Eq.(\ref{son}). Please note that we have subtracted the mean value of the order parameter which only contributes to the $\omega=0$ frequency, which is not of relevance for our analysis.
\begin{figure}[h]
  \includegraphics[width=0.9\columnwidth,clip,angle=0]{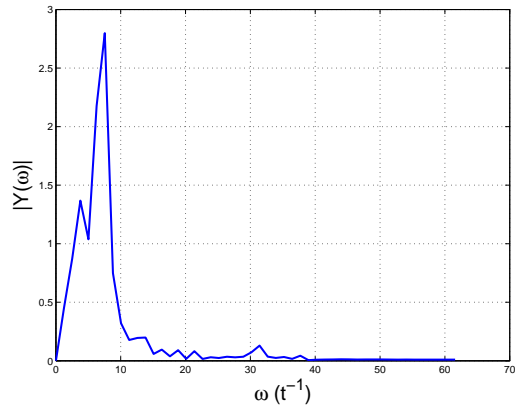}
  \caption[]{(Color online) The frequency spectrum corresponding to the order parameter after the perturbation Eq.\eqref{sina} with $a=0.1, \mu=3$ and $\omega=7$, {\it i.e.}, the top plot in Fig.\ref{fig6}. The first two highest modes are $\omega=3.77$ and $7.54$. One of the frequencies is close to the external driving frequency $\omega=7$. This may indicate that in this case the time evolution of the order parameter is controlled by interference effects between the normal modes and external frequency.}
 \label{fig7}
 \end{figure}
We clearly observe in both figures that for weak quenches ($\tau=1$,$v=0.3$) only very few Fourier modes are excited. However for the stronger quenches ($\tau=0.2$,$v=3$) {we observe multiple peaks and a finite support in a relatively large window of frequencies}. This is a strong suggestion that for strong quenches the time evolution is already quasi-chaotic. For a rigorous demonstration we would have to show that higher order correlation functions, such as $\langle O_2(t) O_2(t')\rangle$, decay exponentially for sufficiently long time separations $|t-t'|$. However this is beyond the scope of the paper.
In Fig.\ref{fig7} we also plot the frequency spectrum corresponding to the time evolution of the order parameter after the oscillatory perturbation Eq.\eqref{sina} with, $a=0.1, \mu=3$ and $\omega=7$.
As it can be observed, the highest mode of the Fourier spectrum is close to the driving frequency $\omega=7$ which indicates that the dynamics of the order parameter is well described by the superposition of the driving frequency and the closest normal modes.
\section{Conclusions}
 We have investigated the time evolution of a holographic superconductor in a soliton background by computing the NM's and the explicit time evolution of the order parameter after a perturbation.
  Full relaxation to equilibrium never occurs in the strongly coupled superconductor that we study. This is a consequence of the lack of horizon in the soliton background. For slow perturbation the order parameter
performs undamped sinusoidal oscillations typical of integrable systems which do not reach thermal equilibrium after a quench.  The frequency of the oscillations is in fair agreement with the lowest NM of the system.
 For sufficiently fast, and strong quenches, many modes contribute to the time evolution of the order parameter. The resulting pattern of oscillations of the order parameter becomes increasingly chaotic with no trace of the single oscillations typical of smooth and slow perturbation. In these chaotic region although a horizon is evidently not formed the order parameter reaches some sort of quasi equilibrium with deviations that superficially resemble noise. \\
 As was mentioned previously the probe limit is enough because the transition is not induced by a dynamical instability \cite{horo3} and because previous results including backreaction \cite{benho} were qualitatively similar. Finally we stress that our results are valid provided that quantum gravity corrections are neglected. For sufficiently long times it is expected that these corrections will induce some sort of thermalization in the system.  However, even if thermalization finally occurs, its typical equilibration time will be much larger than in other field theories whose gravity dual has a horizon\cite{chesler,murata}. \\

% Finally we note that the time evolution after a quench can also be studied in cold atom experiments %as the interaction strength can be controlled by an external magnetic field that tune the distance %between hyperfine levels (Feshbach resonance). The resulting dynamics of the order parameter can be %studied experimentally by radio frequency techniques \cite{rf}.
%   Stronger and faster quenches, related to the existence of relatively close NM's, for which the %oscillations of the order parameter follow an intricate pattern. Since the dual field theory has a mass gap %we speculate that the physical mechanism that prevents equilibration is related to localization caused by %interactions. \\

%Note: At the time we were finalizing this work a paper \cite{basu}
%was posted on arXiv which also studies a quench in an AdS Soliton background.

\vspace{0.45 cm}
AMG thanks Tadashi Takayanagi, Benson Way and especially Norihiro Tanahashi for illuminating conversations. HQZ is delighted to thank Yong-Qiang Wang for his help to understand the code. AMG acknowledge supports from EPSRC, grant No. EP/I004637/1. AMG, HQZ and HBZ acknowledge support from a FCT, grant PTDC/FIS/111348/2009 and a Marie Curie International Reintegration Grant
PIRG07-GA-2010-268172. XG is supported by the MPG-CAS Joint Doctoral Promotion Programme. HBZ is also
partially supported by the National Natural Science Foundation of China (under Grant Nos. 11205020)

% References %%%%%%%%%%%%%%%%%%%%%%%%%%%%%%%%%%%%%%%%%%%%%%%%%%%%%%%%%%%%%%%%%%%%%


\begin{thebibliography}{99}
\bibitem{hai1}We can easily deduce that if we ignore the $r$-component
of the gauge field, the resulting solutions of the EOM will only be
static, {\it i.e.}, independent of time. Therefore, in order to
investigate the dynamical evolution of the system, we must also include
$A_r$.
\bibitem{hai2} The small constant is $\epsilon=0.01$ in $\psi_1(t)=\epsilon+J\tanh(vt)$. The reason for this small constant is that we have chosen Eq.\eqref{eom4} to solve the time evolution of $M_t$ in the Runge-Kutta method. Therefore, in this case one needs to divide by $|\psi|$ on both sides of Eq.\eqref{eom4} in order to satisfy the form of the Runge-Kutta algorithm. However, in this case an individual term $|\psi|$ will appear in the denominator. In order to avoid the numerical divergence of the quench  $\psi_1(t)=J\tanh(vt)$ at $t=0$, one needs to add a small constant term in this quench. In fact one cannot avoid to use Eq.\eqref{eom4} when solving the EoMs by Runge-Kutta method in time direction. Therefore, this is an indispensable trick to obtain meaningful results.
\end{thebibliography}
\end{document}